\documentclass[
aps,
prc,
floatfix,
twocolumn,
superscriptaddress,
amssymb,
amsmath
]{revtex4}

\usepackage{rotating}
\usepackage{mathtools}
\usepackage{tablefootnote}
\usepackage{cleveref}

\newcommand{\alphan}{$(\alpha, \mathrm{n})$ }

\usepackage{graphicx}
\usepackage[usenames]{color} 

\begin{document}

\preprint{APS/123-QED}

\title{Nuclear physics uncertainties in neutrino-driven, neutron-rich supernova ejecta}

\author{J. Bliss} \affiliation{Institut f\"ur Kernphysik, Technische Universit\"at Darmstadt, Schlossgartenstr. 
2, Darmstadt 64289, Germany}


\author{A. Arcones} \affiliation{Institut f\"ur Kernphysik, Technische Universit\"at Darmstadt, Schlossgartenstr. 2, Darmstadt 64289, Germany}
\affiliation{GSI Helmholtzzentrum f\"ur Schwerionenforschung GmbH, Planckstr. 1, Darmstadt 64291, Germany}


\author{F. Montes} \affiliation{National Superconducting Cyclotron
  Laboratory, Michigan State University, East Lansing, MI 48824, USA}
\affiliation{Joint Institute for Nuclear Astrophysics,
  http://www.jinaweb.org}
  

\author{J. Pereira} \affiliation{National Superconducting Cyclotron
  Laboratory, Michigan State University, East Lansing, MI 48824, USA}
\affiliation{Joint Institute for Nuclear Astrophysics,
  http://www.jinaweb.org}
  
 
\begin{abstract}
\textbf{Background:}
Neutrino-driven ejecta in core collapse supernovae (CCSNe) offer an interesting astrophysical scenario where lighter heavy elements between Sr and Ag can be synthesized. Previous studies emphasized the important role that ($\alpha,n$) reactions play in the production of these elements, particularly in neutron-rich and alpha-rich environments. 

\textbf{Purpose:}
In this paper, we have investigated the sensitivity of elemental abundances to specific ($\alpha,n$) reaction-rate uncertainties under different astrophysical conditions. 

\textbf{Method:}
The abundances of ligther heavy elements were calculated with a reaction network under different astrophysical conditions. ($\alpha,n$) reaction rates were varied within their theoretical uncertainty using a Monte Carlo approach.

\textbf{Results:}
The most important ($\alpha,n$) reaction affecting the nucleosynthesis of lighter heavy nuclei were identified for 36 representative conditions of CCSNe neutrino-drive winds.

\textbf{Conclusions:}
Experimental studies of these reactions will reduce the nucleosynthesis uncertainties and make it possible to use observations to understand the origin of lighter heavy elements and the astrophysical conditions where they are formed.
\end{abstract}

\maketitle


\section{Introduction}
\label{sec:introduction}

Until relatively recently, the origin of elements heavier than iron was thought to be limited to the s-process (weak and strong) and  the r-process, with only a few isotopes been produced by the p-process. We know now that several other processes may produce elements beyond iron (e.g., $\nu$p-process \citep{Froehlich.etal:2006, Wanajo:2006, Pruet.etal:2006}, i-process \citep{Cowan1977}). Observations provide a unique window to look into the origin of elements. By combining nucleosynthesis studies based on different astrophysical conditions with observations, one can learn about the environments where heavy elements are synthesised \cite[see e.g.][]{Hansen.etal:2014}.

Enormous progress has been reported in our understanding of the r-process in the last years \citep{Horowitz.etal:2019}. The kilonova associated with the gravitational wave detection, GW170817 \citep{Cowperthwaite.etal:2017, Abbott.etal:2017} and by the first identification of a heavy element, strontium, in its spectrum \citep{Watson.Hansen.etal:2019} are indications that production of heavy elements occurs in neutron star mergers. Furthermore, since the r-process conditions found in such mergers may be different in different parts of the ejecta, the change of the kilonova from blue to red may indicate that there may not be a unique r-process abundance pattern.  Additional evidence for multiple r-process conditions are found in the observations of oldest stars in our galaxy and dwarf galaxies \citep{Mashonkina2017, skuladottir2019, Reichert.etal:2019} and in recent results from galactic chemical evolution models \cite[e.g.,][]{Cote.etal:2019}. The scattered abundances of elements below the second r-process peak (from Sr to Ag), observed in different Eu-enriched stars, contrasts with the rather robust pattern found for elements between the second and third r-process peak. Moreover, some of the stars present high abundances of lighter heavy elements (defined here as elements from Sr to Ag) compared to elements beyond the second r-process peak (these stars have been referred as Honda-like stars \citep{Honda.etal:2004} or r-limited stars \citep{THansen.etal:2018}). Already \cite{Travaglio.etal:2004, Montes.etal:2007, Qian.Wasserburg:2002} demonstrated that at least one additional process is necessary to explain the observed solar and old-star abundances.  

One possible origin of these lighter heavy elements are the neutrino-driven ejecta in core-collapse supernovae. At the end of their life, massive stars collapse, form a neutron star, and a shock wave is launched and destroys the star. The details of the explosion have not been fully understood but there is a consensus that standard supernovae are driven by neutrinos and hydrodynamical instabilities \citep{Janka:2012, Pan.etal:2019, Just.etal:2018}. A subset of supernovae may be instead triggered by fast rotation and magnetic fields, i.e., magneto-rotational supernova (MR-SN). Nevertheless, in both kind of explosions, part of the matter is ejected by the emitted neutrinos and can become neutron or proton-rich. For MR-SN the magnetic pressure facilitate the ejection of neutron-rich matter and the r-process can produce the heaviest elements (\cite{Winteler.etal:2012, Nishimura.etal:2017, Mosta.etal:2018}). In contrast, in the matter that is mostly ejected by neutrinos (for both types of explosions), neutrino interactions affect the proton-neutron composition leading to conditions favourable to produce lighter heavy elements up to the second r-process peak or below. Current simulations indicate that significant part of the ejecta is proton rich \citep{Wanajo.etal:2018} and that it is possible to have small amounts of fast-expanding, neutron-rich material \cite[see, e.g.][]{Wanajo.etal:2011}. In this paper,  we focus on the nucleosynthesis occurring in slightly neutron rich ejected material.

The nucleosynthesis of this neutrino-driven, neutron-rich ejecta has been investigated by several groups \cite{Hoffman.etal:1996, Arcones.Bliss:2014, Arcones.Thielemann:2013}. The energy deposited by neutrinos leads to unbound matter that expands and cools. Initially the temperature of the ejected matter is high and thus the composition is dominated by neutrons and protons. During the expansion, the temperature drops and $\alpha$-particles form and recombine into seed nuclei still in nuclear statistical equilibrium (NSE). Due to the fast expansion, the temperature further drops and an alpha-rich NSE freeze-out occurs at $T\sim5$~GK. If the neutron-to-seed and alpha-to-seed ratios are relative high \cite{Bliss.etal:2018}, then matter can reach heavier nuclei mainly by ($\alpha$,n) reactions. This is known as alpha process, charged-particle reaction (CPR) phase, or weak r-process and lasts until the temperature drops below $T\sim 2$~GK. During this phase, there is an (n,$\gamma$)-($\gamma$,n) equilibrium in every isotopic chain with the maximum abundances at few isotopes away from stability. Furthermore, since the expansion time scale is relatively short (tens of mili-seconds),  the beta decays are too slow compared to ($\alpha$,n) reactions that have dropped out of equilibrium with their (n,$\alpha$) counterpart. Those reactions become the main channel to move matter towards heavier nuclei, with a minor contribution from (p,n) and ($\alpha$,$\gamma$) reactions. In order to fully understand this process, uncertainties in both the specific astrophysical conditions in the wind, and nuclear physics uncertainties in the reactions involved have to be quantified and reduced.

Our aim in this paper is to identify the key reactions that need to be measured to reduce the nuclear physics uncertainty to be able to use observations to constrain astrophysical wind conditions. In Bliss 2017, we studied the overall effect of both (astrophysical conditions and nuclear physics uncertainties) and concluded that  ($\alpha$,n) reactions are critical and still not well constrained from theory or experiments \citep{Pereira.Montes:2016,Mohr:2016}. In a follow-up study \citep{Bliss.etal:2018}, a systematic nucleosynthesis study covering all possible astrophysical wind conditions was performed. In this paper, we explore the impact of ($\alpha$,n) reactions using a Monte Carlo study for 36 representative trajectories covering a broad range of astrophysical conditions. We provide for the first time a list of key reactions for the weak r-process in core-collapse supernovae by investigating correlations, impact on the abundances, and importance under different astrophysical conditions.

The paper is structured as follows.  We discuss the astrophysical conditions of the 36 trajectories selected in Sect.~\ref{sec:astro}. The Monte Carlo method is introduced in Sect.~\ref{Sect.:MCapproach}, including the identification of key reactions. In Sect.~\ref{Sect.:Results}, we present our results including the list of key reactions. We conclude in Sect.~\ref{sec:conclusions}.

\section{Variety in the astrophysical conditions in neutrino-driven supernova ejecta}
\label{sec:astro}
In core-collapse supernovae, matter can be shock-heated and become unbound or it can be ejected by neutrinos. In the neutrino-driven ejecta, neutrinos deposit enough energy to unbound  matter while they also change neutrons into protons. Therefore, the properties of the neutrino-driven ejecta can vary from neutron- to proton-rich and produce elements up to around Silver \cite[see e.g.,][for a review]{Arcones.Bliss:2014}. When matter expands very fast, neutrinos cannot act long enough and the ejected matter stays neutron rich. In slightly neutron-rich conditions (electron fractions $Y_{\mathrm{e}}$ between 0.4 and 0.5) the nucleosynthesis path runs close to stability on the neutron-rich side or along the valley of stability. 

\begin{figure}[h!]
\centering
\includegraphics[width=1.0\linewidth,angle=0]{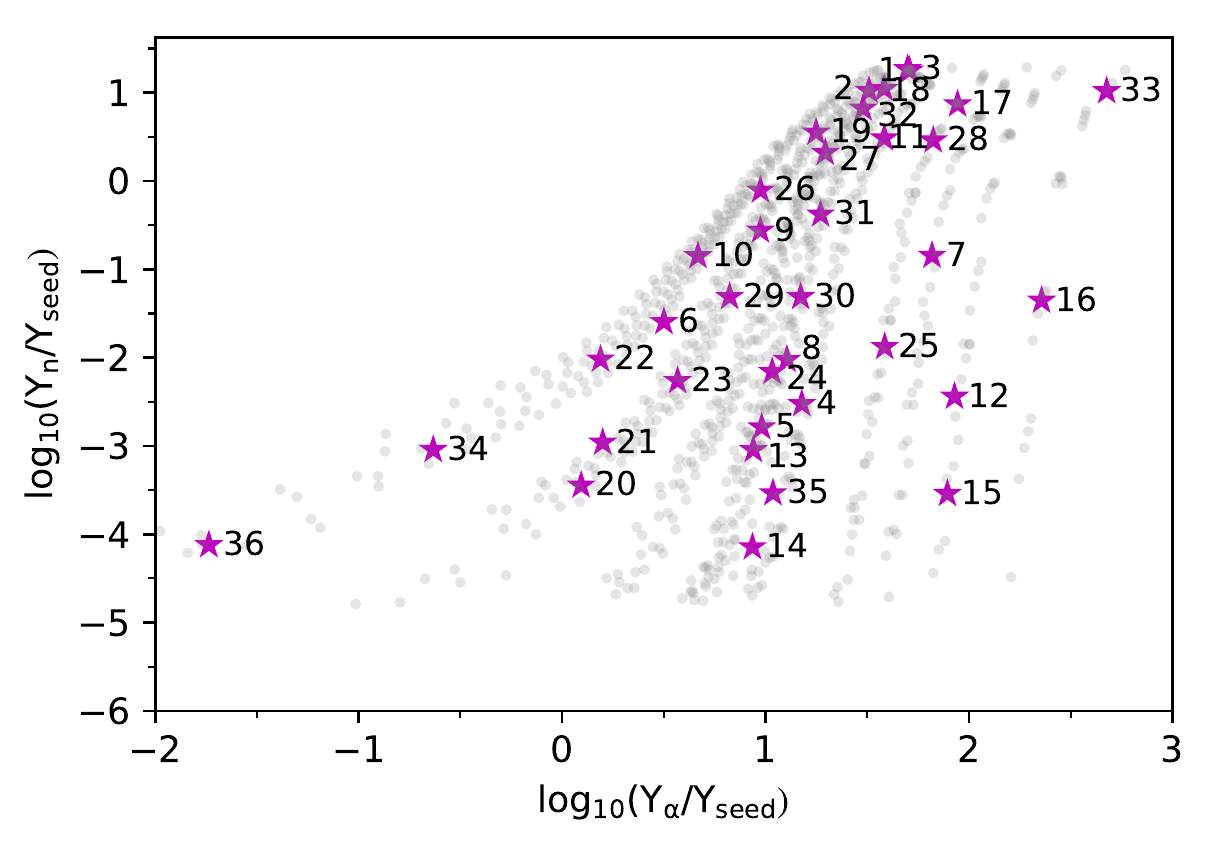}
\caption{Distribution of the CPR2 tracers from \citet{Bliss.etal:2018} in the $Y_{\alpha}/Y_{\mathrm{seed}}$-$Y_{n}/Y_{\mathrm{seed}}$ plane at 3~GK. The stars mark the astrophysical conditions for which we performed sensitivity studies.}
\label{fig:YnYalphaYseed}
\end{figure}

In \cite{Bliss.etal:2018}, we have performed a systematic study covering a broad range of possible astrophysical conditions. Our study was based on a steady-state model for neutrino-driven winds (following \cite{Otsuki.etal:2000}) using, as input parameters, the proto-neutron star masses and radii, neutrino luminosities and energies, and initial electron fraction ($Y_{\mathrm{e}}$). This allowed us to investigate the sensitivity of the weak r-process to the wind parameters (i.e., $Y_{\mathrm{e}}$, entropy, expansion timescale). The conclusion of that paper was that the final abundances can be divided into four distinctive patterns (referred to as NSE1, NSE2, CPR1, and CPR2, see Fig.~4 in \cite{Bliss.etal:2018}) and that each of these abundance-pattern groups are clearly correlated with the neutron-to-seed ($Y_{\mathrm{n}}/Y_{\mathrm{seed}}$) and $\alpha$-to-seed ($Y_{\alpha}/Y_{\mathrm{seed}}$) ratios at $T=3$~GK. The NSE1 and NSE2 abundance distributions are mainly set already during the nuclear statistical equilibrium (NSE) phase. Therefore, binding energies and partition functions of the involved nuclei, and not individual reactions rates, determine the nucleosynthesis (see also \citet{Wanajo.etal:2018} for a similar conclusion). For wind conditions resulting in a CPR1 pattern, the final abundances are mainly given by known Q-values of $(\alpha,\mathrm{n})$ reactions at $N=50$, as also concluded by \cite{Hoffman.etal:1996,Wanajo:2006}. The group CPR2 is linked to high $Y_{\mathrm{n}}/Y_{\mathrm{seed}}$ and $Y_{\alpha}/Y_{\mathrm{seed}}$. For this group, individual nuclear reaction rates have a critical impact on the nucleosynthesis evolution and on the final abundances. Moreover, the elemental abundance patterns, which can extend up to $Z=55$, are highly dependent on the specific wind conditions. Therefore, in order to identify the most important reactions we have to explore several astrophysical conditions that lead to different abundances within CPR2. The conditions for the group CPR2 are shown in Fig.~\ref{fig:YnYalphaYseed} with stars indicating the selected trajectories for our sensitivity study.  Table~\ref{tab:MCs} provides the  wind parameters for the 36 trajectories.

\begin{table}[!htb]
 \begin{center}
  \caption{Astrophysical conditions associated with each trajectory }
  \label{tab:MCs}
  \vspace*{1mm}
  \begin{tabular}{cccc}
  \hline
  \hline
Trajectory  & $Y_{e}$ & Entropy & Expansion time  \\
& &$k_{\mathrm{B}}/\mathrm{nuc}$ & ms \\
  \hline
MC1 & 0.42 & 129 & 11.7 \\
MC2 & 0.45 & 113 & 11.9 \\
MC3 &  0.45 & 122 & 10.3 \\
MC4 &  0.44 & 66 & 19.2 \\
MC5 &  0.43 & 66 & 34.3 \\
MC6 &  0.4 & 56 & 63.8 \\
MC7 &  0.47 & 96 & 11.6 \\
MC8 &  0.43 & 78 & 35 \\
MC9 & 0.40 & 73 & 28.1 \\
MC10 &  0.40 & 54 & 31 \\
MC11 &  0.44 & 104 & 13.2 \\
MC12 & 0.48 & 85 & 9.7 \\
MC13 &  0.43 & 64 & 35.9 \\
MC14 &  0.45 & 46 & 14.4 \\
MC15 & 0.48 & 103 & 20.4 \\
MC16 &  0.49 & 126 & 15.4 \\
MC17 &  0.46 & 132 & 12.4 \\
MC18 &  0.45 & 131 & 21.4 \\
MC19 &  0.41 & 75 & 9.8 \\
MC20 &  0.41 & 42 & 59.3 \\
MC21 &  0.41 & 31 & 22.2 \\
MC22 &  0.40 & 40 & 46.7 \\
MC23 &  0.41 & 48 & 37.5 \\
MC24 &  0.43 & 56 & 16.2 \\
MC25 &  0.46 & 96 & 20.9 \\
MC26 &  0.40 & 84 & 36.2 \\
MC27 &  0.42 & 76 & 10 \\
MC28 & 0.46 & 113 & 11.9 \\
MC29 &  0.41 & 66 & 41.4 \\
MC30 & 0.43 & 79 & 26.3 \\
MC31 &  0.43 & 71 & 11.4 \\
MC32 &  0.42 & 103 & 12.7 \\
MC33 &  0.49 & 175 & 14.2 \\
MC34 &  0.40 & 34 & 58.7 \\
MC35 &  0.44 & 48 & 13 \\
MC36 &  0.40 & 32 & 63.4 \\
  \hline
  \end{tabular}
 \end{center}
\end{table}

\section{Sensitivity study on $(\alpha, \mathrm{n})$ reaction rate uncertainties} 
\label{Sect.:MCapproach}
For the 36 trajectories introduced before, we performed a Monte Carlo study to investigate the impact of $(\alpha, \mathrm{n})$ reactions. A similar approach was used in other sensitivity studies  for Type I X-ray bursts~\citep{Parikh.etal:2008}, novae ~\citep{Smith.etal:2002,Hix.etal:2003}, p-process~\citep{Rauscher.etal:2016}, s-process~\citep{Nishimura.etal:2017,Cescutti.etal:2018}, r-process nucleosynthesis~\citep{Mumpower.etal:2016,Mumpower.etal:2017}, and $\nu$p-process~\citep{Nishimura.etal:2019}. 

We considered 909 $(\alpha, \mathrm{n})$ reactions on stable and neutron-rich nuclei between Fe (Z=26) and Rh (Z=45). The reference (or baseline) $(\alpha, \mathrm{n})$ reaction rates were calculated with TALYS~1.6 using the nuclear physics referred as TALYS~1 in \citet{Pereira.Montes:2016} (see their Table II) except for the masses, which were taken from \citet{Audi.etal:2003} if available, or from the FRDM mass model \citep{Moeller.etal:1995} otherwise. 

For each of the 36  trajectories considered, more than 10000 network calculations were performed each with different $(\alpha, \mathrm{1n})$ reaction rates. Only $(\alpha, \mathrm{1n})$ reactions rates were varied since, within the relevant temperature range (1 GK $\lesssim T \lesssim$ 5 GK), the 1-neutron emission channel usually dominates over the emission of multiple neutrons with very few exceptions at high temperatures. Even in the cases where 2-neutron channel is comparable to the 1-neutron channel, the assumption is justified since at those temperatures, the rapid establishment of an equilibrium isotopic distribution by $(n,\gamma)$ and $(\gamma,n)$ reactions makes the creation of the heavier $Z+2$ nucleus and corresponding neutron emission(s) independent of the particular $(\alpha,\mathrm{xn})$ production reaction channel(s). In each network calculation, each of the 909 $(\alpha, \mathrm{1n})$ reaction rates were varied simultaneously within the expected uncertainties by independently applying a randomly-distributed scaling factor $p$. The same factor was applied to the corresponding forward and reverse rates. The rate variation factors $p$ was chosen to follow a log-normal distribution (if $p$ is log-normally distributed, $\ln p$ is normally distributed). Because the log-normal density function is only defined for $p \geq 0$, the rate variation factors are always positive. Since the average theoretically-estimated uncertainty of the $(\alpha, \mathrm{n})$ reaction rates is about a factor 10 within the temperatures of interest \citep{Pereira.Montes:2016,Bliss.etal:2017} the mean value and the standard deviation of $\ln p$ were $\mu = 0$ and $\sigma = 2.3$, respectively (corresponding to having a variation factor between
0.1 and 10 with a probability of 68.3\%).

\subsection{Identification of key $(\alpha, \mathrm{n})$ reactions}
\label{Sect.:Correlation}
The most influential reaction rates on the final abundances were identified by calculating correlation coefficients between the variations of the rates and the resulting abundance changes. The correlations were determined using the Spearman's correlation coefficient \citep{Spearman:1904}. The Spearman's rank-order correlation coefficient $r_{\mathrm{corr}}$ measures the strength and direction of the monotonic relationship between two variables (i.e., rate variation factor $p$ and elemental abundance $Y$) using their ranks. In case of a monotonic relationship the value of one variable either increases or decreases as the other value increases. Previous sensitivity studies (e.g., \citet{Rauscher.etal:2016,Nishimura.etal:2017}) applied the Pearson correlation coefficient \citep{Pearson} which quantifies the strength of the linear relationship between two variables. Since nucleosynthesis calculations frequently show non-linear relations between variations of reaction rates and abundance changes (see, e.g., Fig.~6 \citet{Iliadis.etal:2015}) we rely on the Spearman's $r_{\mathrm{corr}}$ which is better suited to deal with non-linear behavior.

The Spearman's correlation coefficient is calculated using  
\begin{equation}
r_{\mathrm{corr}} = \frac{\sum_{i=1}^{n}\left(R(p_{i}) - \overline{R(p)}\right)\left(R(Y_{i}) - \overline{R(Y)}\right)}{\sqrt{\sum_{i=1}^{n}\left(R(p_{i}) - \overline{R(p)}\right)^{2}}\sqrt{\sum_{i=1}^{n}\left(R(Y_{i}) - \overline{R(Y)}\right)^{2}}},
\label{Eq.:CorrCoeff}
\end{equation}
where $n$ is the number of network calculations for a given trajectory, $R$ corresponds to the ranks of the rate variation factors $\lbrace p_{1},p_{2},...,p_{n}\rbrace$ and final abundances $\lbrace Y_{1},Y_{2},...,Y_{n}\rbrace$, and $\overline{R(p)}=\left(\sum_{i=1}^{n}R(p_{i})\right)/n$ and $\overline{R(Y)}=\left(\sum_{i=1}^{n}R(Y_{i})\right)/n$ are the average ranks.  The values of $r_{\mathrm{corr}}$ range between $-1 \leq r_{\mathrm{corr}} \leq +1$. A Spearman's correlation factor of $+1$ ($-1$) indicates a perfectly increasing (decreasing) monotonic function.

\begin{figure*}[htb!]
\centering
\includegraphics[width=1.0\linewidth,angle=0]{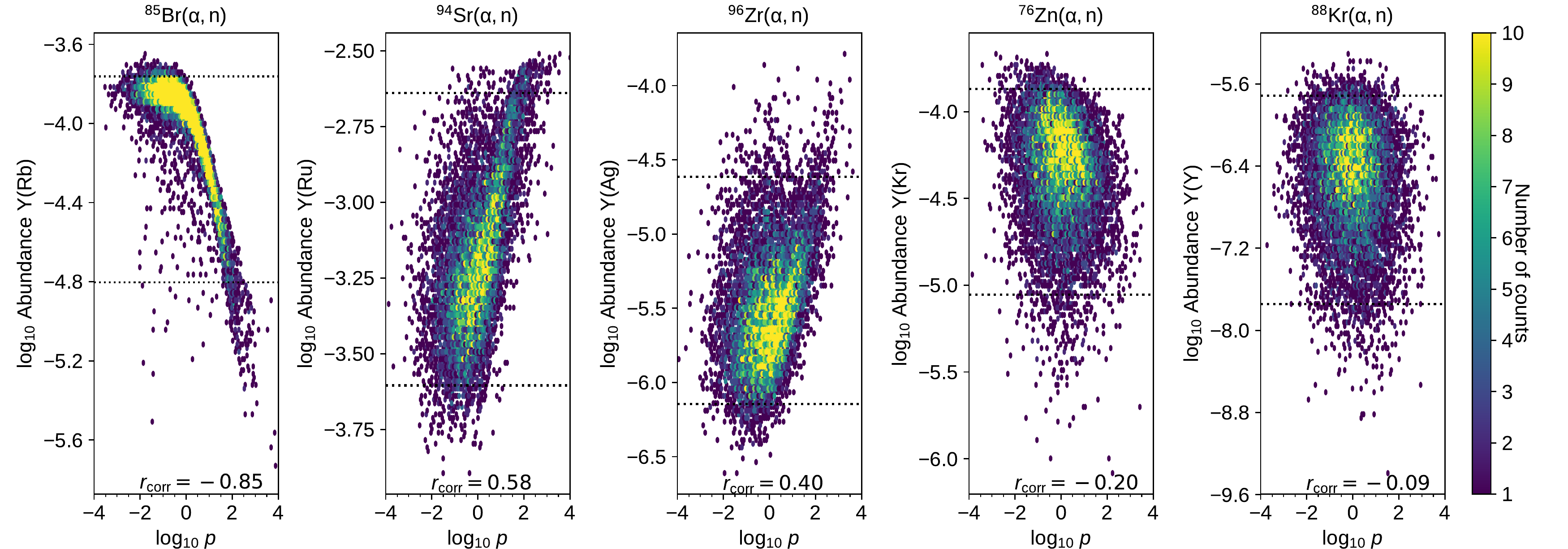}
\caption{Impact of the variation of the $^{85}\mathrm{Br}(\alpha,n)$, $^{94}\mathrm{Sr}(\alpha,n)$, $^{96}\mathrm{Zr}(\alpha,n)$, $^{76}\mathrm{Zn}(\alpha,n)$, and $^{88}\mathrm{Kr}(\alpha,n)$ reaction rates on the abundances of Rb (first panel), Ru (second panel), Ag (third panel), Kr (fourth panel) and Y (fifth panel), respectively. The dashed lines illustrate distribution of the abundances in 95.4\% of the network runs The color code denotes how often the same abundance occurs for the same rate variation factor. There is a very strong negative correlation between $^{85}\mathrm{Br}(\alpha,n)$ and the Rb abundances, a strong positive correlation between $^{94}\mathrm{Sr}(\alpha,n)$ and the Ru abundances, and no correlation between $^{88}\mathrm{Kr}$ and the abundance of Y.}
\label{fig:Spearman}
\end{figure*}

Figure~\ref{fig:Spearman} shows several cluster-plots with elemental abundances as a function of the logarithm of the scaling factor $p$ for different ($\alpha,n$) reactions. The calculations were done using the trajectory MC27 (see Table~\ref{tab:MCs}). These cluster-plots illustrate the correlations between the abundances of Rb and $^{85}\mathrm{Br}(\alpha,\mathrm{n})$ ($r_{\mathrm{corr}}=-0.85$); Ru and $^{94}\mathrm{Sr}(\alpha,\mathrm{n})$ ($r_{\mathrm{corr}}=0.58$); Ag and $^{96}\mathrm{Zr}(\alpha,\mathrm{n})$ ($r_{\mathrm{corr}}=0.40$); Kr and $^{76}\mathrm{Zn}(\alpha,\mathrm{n})$ ($r_{\mathrm{corr}}=-0.20$); and Y and $^{88}\mathrm{Kr}(\alpha,\mathrm{n})$ ($r_{\mathrm{corr}}=-0.09$). Each data point corresponds to a single reaction network calculation where all reaction rates were varied simultaneously. Data points within the dashed lines indicate nucleosynthesis calculations resulting in an elemental abundances within $\pm$2-$\sigma$ of the average abundance. Figure~\ref{fig:Spearman} shows a very strong negative correlation between the variation of $^{85}\mathrm{Br}(\alpha,\mathrm{n})$ and the Rb abundance, especially for $\log _{10} p > 0$. In comparison the correlation between $^{94}\mathrm{Sr}(\alpha,\mathrm{n})$ and the abundance of Ru is positive. The absolute value of the correlation factor for a given element and reaction will be smaller the more reactions contribute to the abundance uncertainty of that element (see e.g the influence of $^{96}\mathrm{Zr}(\alpha,\mathrm{n})$ and $^{88}\mathrm{Kr}(\alpha,\mathrm{n})$ on the abundance of Kr and Y, respectively). If $r_{\mathrm{corr}}$ is close to zero, as for Yttrium abundance and $^{88}\mathrm{Kr}(\alpha,\mathrm{n})$, there is no significant correlation between the rate variation factors and the final abundances. We have used a correlation factor value $|r_{\mathrm{corr}}| \geq 0.20$ as the threshold to indicate a meaningful correlation between a specific ($\alpha,n$) reaction rate and an elemental abundance change. It should be noted that, for a given element, the correlation factor alone should not be used to rank the important reactions since the Spearman's correlation coefficient is independent of the magnitude of the element abundance variation (i.e., the slope of the cluster). We therefore identified important key reactions by 1) inspecting which abundances vary most in the Monte Carlo study (large slopes); 2) investigating which reaction rate variations strongly correlate with absolute abundance changes (large $r_{\mathrm{corr}}$).  We focused our study on the aforementioned lighter r-elements $Z=36-47$.

\section{Results} 
\label{Sect.:Results}
We have identified 45 \alphan reactions having an impact on the elemental abundances. Tables~\ref{tab:Z36}--\ref{tab:Z47} show the \alphan reactions which affect the elemental abundances between $Z=36-47$ and satisfy the following conditions for at least one of the MC1-C36 wind trajectories: 
\begin{itemize}
	\item $|r_{\mathrm{corr}}| \geq 0.20$ 
	\item An abundance variation at least a factor of 5 between the maximum and minimum elemental abundance within $2-\sigma$ of the average abundance
\end{itemize}

All of the important reactions can be classified in three categories depending on whether they involve nuclei with $N<50$, $N=50$ or at $N>50$. The $N=50$ shell closure serves as a process bottleneck at temperatures around 4-5~GK due to the $(n,\gamma)-(\gamma,n)$ equilibrium. $N=50$ Isotopes with the highest abundances at those temperatures ($^{80}\mathrm{Zn}$, $^{81}\mathrm{Ga}$, $^{82}\mathrm{Ge}$,  $^{83}\mathrm{As}$, $^{84}\mathrm{Se}$, $^{85}\mathrm{Br}$, $^{86}\mathrm{Kr}$, $^{87}\mathrm{Rb}$) determine the speed at which \alphan reactions move the material flow towards heavier nuclei. There are two ways reactions on shell closure isotopes affect the nucleosynthesis. In some cases, the abundance flow is stopped or slowed down at the $N=50$ shell closure once the temperatures drops below 2~GK. In these cases, the final abundances are established by the $\beta$-decay to stability of those isotopes. An example is the Kr abundance (see Tab.~\ref{tab:Z36}) for which  the largest impact is directly due the reactions on $^{84,85}\mathrm{Se}$. Similar cases are the abundances of Rb and Sr which are directly affected by reactions on $^{83}\mathrm{As}$, $^{84,85}\mathrm{Se}$, $^{85}\mathrm{Br}$ and $^{86}\mathrm{Kr}$ (for Sr). The effect of the $N=50$ closure is not only due to the direct effect of bottle-necked nuclei $\beta$-decaying to stability but also due to the indirect effect of hindering (or enhancing) the  $N>50$ abundances. Therefore reactions involving $N=50$ nuclei also affect the final abundances of Y, Zr, Nb, Mo, Ru and Rh. For the heavier elements, the importance of the $N=50$ closure is diluted by \alphan reactions on $N>50$ nuclei. For a large number of conditions, \alphan reactions on $^{84,85}\mathrm{Se}$ direct the flow out of the $N=50$ shell closure. Therefore those reactions directly affect a large number of abundances. 

Since the system is in $(n,\gamma)-(\gamma,n)$ equilibrium, reactions on the most abundant isotopes of a given isotopic chain can also indirectly affect final abundances. For example, reactions on $^{88,90}\mathrm{Kr}$ affect the Rb abundance even though the heaviest stable rubidium isotope is $^{87}\mathrm{Rb}$ (produced by the beta decay of $^{87}\mathrm{Kr}$).  
The $^{88,90}\mathrm{Kr}$ \alphan reactions indirectly affect the amount of $^{87}\mathrm{Kr}$ by modifying the overall Kr isotopic abundances. Almost all of the reactions affecting elements Ru, Rh, Pd and Ag are of this type. There are a few reactions that affect final abundances due to their effect in the neutron abundance (once the material is running out of neutrons). For example, the  $^{94}\mathrm{Sr}(\alpha,\mathrm{n})$ affects the Rb abundance due to the change in neutron abundance at late times.

Reactions on $N<50$ isotopes have an impact only for a limited set of similar astrophysical conditions (see Table~\ref{tab:complete} and corresponding tracers in Fig.~\ref{fig:YnYalphaYseed}). An example of this are the reactions on $^{74, 76}\mathrm{Ni}$ and $^{76, 78, 79, 80}\mathrm{Zn}$ affecting the final Kr abundance.
Since for some conditions, $^{78,79}\mathrm{Zn}$ are the entry gateways to the $N=50$ shell closure, \alphan reactions on them influence Kr, Rb, Sr, Y, Zr, Nb and Mo abundances. In general, it is observed that $N<50$ \alphan reactions lead to  smaller abundance variation and correlation coefficients $|r_{\mathrm{corr}}|$ than $N=50$ and $N>50$ reactions. This indicates that in addition to those \alphan reactions other reactions contribute, to a lesser degree, to the change of the final abundance. 

In Tab.~\ref{tab:complete}, we give a complete overview of all \alphan reactions having an influence on the elemental abundances between $Z=36-45$ for at least one MC tracer
Reactions can be grouped according to how many final abundances are effected and for how many astrophysical conditions:
\begin{itemize}
	\item Many elemental abundances under many astrophysical conditions: $^{84,85}\mathrm{Se}$, $^{86-90}\mathrm{Kr}$, $^{90-94}\mathrm{Sr}$, $^{96}\mathrm{Zr}$
	\item Few elemental abundances under many astrophysical conditions: $^{85,87}\mathrm{Br}$, $^{94}\mathrm{Y}$, $^{97,98}\mathrm{Zr}$
	\item Many elemental abundances under few astrophysical conditions: $^{59}\mathrm{Fe}$, $^{63}\mathrm{Co}$, $^{74,76}\mathrm{Ni}$, $^{72,76, 78-80}\mathrm{Zn}$, $^{83}\mathrm{Ge}$, $^{81}\mathrm{Ga}$, $^{78,80,82}\mathrm{Ge}$
	\item Few elemental abundances under few astrophysical conditions: $^{68}\mathrm{Fe}$, $^{71}\mathrm{Co}$, $^{67,77}\mathrm{Cu}$, $^{88}\mathrm{Br}$, $^{88-90}\mathrm{Sr}$, $^{95}\mathrm{Y}$, $^{94,95}\mathrm{Zn}$, $^{97}\mathrm{Nb}$
\end{itemize}
The reduction of the uncertainties in the named reaction rates will contribute to better understand the formation of the lighter heavy elements and help constrain on the astrophysical conditions where they are synthesized.     

\begin{table*}[!htb]
 \begin{center}
  \caption{Element ($Z$) and wind trajectories for which the Spearman's coefficient satisfies $|r_{\mathrm{corr}}| \geq 0.20$ and the elemental abundance varies by at least a factor of 5 within $2-\sigma$ of the abundance distribution. See text for details. The tracer MC1 does not appear in the table because \alphan reactions are not important if $Y_n/Y_{\mathrm{seed}}$ is large and the nucleosynthesis path evolves further away from stability. The MC1 tracer has only an impact on $Z=55$ which is above the heaviest abundances we consider here.}
  \label{tab:complete}
  \vspace*{3mm}
  \begin{tabular}{ccc}
  \hline
  \hline
Reaction  & Z & MC tracers  \\
  \hline
$^{59}$Fe$(\alpha,\mathrm{n})^{62}$Ni & 39$-$42, 45 & 34, 36 \\
$^{68}$Fe$(\alpha,\mathrm{n})^{71}$Ni & 36, 37 & 3 \\
$^{63}$Co$(\alpha,\mathrm{n})^{66}$Cu & 39$-$42, 45 & 20, 34, 36 \\
$^{71}$Co$(\alpha,\mathrm{n})^{74}$Cu & 36, 37 & 3 \\
$^{74}$Ni$(\alpha,\mathrm{n})^{77}$Zn & 36$-$42 & 2, 3, 17, 18, 32 \\
$^{76}$Ni$(\alpha,\mathrm{n})^{79}$Zn & 36$-$42 & 2, 3, 18, 32 \\
$^{67}$Cu$(\alpha,\mathrm{n})^{70}$Ga & 47 & 35 \\
$^{77}$Cu$(\alpha,\mathrm{n})^{80}$Ga & 37 & 3 \\
$^{72}$Zn$(\alpha,\mathrm{n})^{75}$Ge & 39$-$42 & 36 \\
$^{76}$Zn$(\alpha,\mathrm{n})^{79}$Ge & 36, 37$-$42 & 2, 3, 17, 18, 32 \\
$^{78}$Zn$(\alpha,\mathrm{n})^{81}$Ge & 36, 37$-$42 & 2, 3, 17, 18, 32 \\
$^{79}$Zn$(\alpha,\mathrm{n})^{82}$Ge & 36, 37$-$42 & 2, 3, 18, 32 \\
$^{80}$Zn$(\alpha,\mathrm{n})^{83}$Ge & 36, 37, 39$-$42 & 2, 3, 18, 32 \\
$^{81}$Ga$(\alpha,\mathrm{n})^{84}$As & 36, 38, 39, 41 & 17, 32 \\
$^{78}$Ge$(\alpha,\mathrm{n})^{81}$Se & 39$-$42 & 36 \\
$^{80}$Ge$(\alpha,\mathrm{n})^{83}$Se & 36$-$39, 42 & 28, 33, 36 \\
$^{82}$Ge$(\alpha,\mathrm{n})^{85}$Se & 36$-$39, 41 & 11, 17, 19, 27, 28, 33 \\
$^{83}$As$(\alpha,\mathrm{n})^{86}$Br & 36, 37, 41 & 11, 26, 27, 28, 33 \\
$^{84}$Se$(\alpha,\mathrm{n})^{87}$Kr & 36$-$42, 44, 45 & 2, 6, 7, 8, 9, 10, 11, 18, 19, 20, 22, 23, 24, 26, 27, 28, 29, 30, 31, 33, 34, 36 \\
$^{85}$Se$(\alpha,\mathrm{n})^{88}$Kr & 36$-$42, 44, 45 & 2, 6, 7, 8, 9, 10, 11, 18, 19, 22, 23, 24, 26, 27, 28, 29, 30, 31 \\
$^{85}$Br$(\alpha,\mathrm{n})^{88}$Rb & 37$-$39 & 6, 7, 8, 9, 10, 22, 23, 24, 26, 28, 29, 30, 31 \\
$^{87}$Br$(\alpha,\mathrm{n})^{90}$Rb & 37, 39 & 6, 9, 10, 29, 31 \\
$^{88}$Br$(\alpha,\mathrm{n})^{91}$Rb & 39 & 26 \\
$^{86}$Kr$(\alpha,\mathrm{n})^{89}$Sr & 38$-$42, 44, 45, 47 & 4, 5, 7, 8, 13, 14, 15, 16, 20, 24, 25, 33, 34, 35 \\
$^{87}$Kr$(\alpha,\mathrm{n})^{90}$Sr & 38$-$42, 45 & 4, 5, 7, 8, 13, 16, 20, 24, 25, 28, 30, 33, 34 \\
$^{88}$Kr$(\alpha,\mathrm{n})^{91}$Sr & 37$-$42, 44, 45 & 2, 4, 5, 6, 7, 8, 9, 11, 13, 16, 17, 18, 19, 20, 22, 24, 25, 26, 27, 28, 29, 30, 31, 32, 33, 34 \\
$^{89}$Kr$(\alpha,\mathrm{n})^{92}$Sr & 39, 40, 42, 44, 45 & 2, 6, 11, 17, 18, 19, 26, 27, 28, 29, 30, 32 \\
$^{90}$Kr$(\alpha,\mathrm{n})^{93}$Sr & 37$-$42, 44$-$46 & 2, 3, 6, 9, 10, 11, 17, 18, 19, 22, 26, 27, 28, 29, 30, 31, 32 \\
$^{87}$Rb$(\alpha,\mathrm{n})^{90}$Y & 41, 45 & 14, 15 \\
$^{89}$Rb$(\alpha,\mathrm{n})^{92}$Y & 41, 42 & 5, 7, 13, 20, 34 \\
$^{88}$Sr$(\alpha,\mathrm{n})^{91}$Zr & 42, 44 & 14, 15 \\
$^{89}$Sr$(\alpha,\mathrm{n})^{92}$Zr & 42 & 14, 15 \\
$^{90}$Sr$(\alpha,\mathrm{n})^{93}$Zr & 42, 44$-$47 & 4, 5, 12, 13, 14, 15, 16, 20, 35 \\
$^{91}$Sr$(\alpha,\mathrm{n})^{94}$Zr & 44, 45 & 5, 12, 13, 16 \\
$^{92}$Sr$(\alpha,\mathrm{n})^{95}$Zr & 38, 42, 44$-$47 & 4, 5, 6, 7, 8, 11, 12, 13, 16, 20, 21, 22, 23, 24, 25, 28, 29, 30, 31, 34 \\
$^{93}$Sr$(\alpha,\mathrm{n})^{96}$Zr & 42, 44$-$47 & 6, 7, 9, 10, 11, 22, 26, 27, 28, 29, 30, 31 \\
$^{94}$Sr$(\alpha,\mathrm{n})^{97}$Zr & 37$-$42, 44$-$47 & 2, 6, 7, 8, 9, 10, 11, 18, 19, 22, 23, 24, 25, 26, 27, 28, 29, 30, 31, 32 \\
$^{94}$Y$(\alpha,\mathrm{n})^{97}$Nb  & 45 & 4, 8, 16, 21, 23, 24, 25 \\
$^{95}$Y$(\alpha,\mathrm{n})^{98}$Nb  & 45, 46 & 8, 23, 24, 25, 30 \\
$^{94}$Zr$(\alpha,\mathrm{n})^{97}$Mo & 44, 45 & 14, 15, 35 \\
$^{95}$Zr$(\alpha,\mathrm{n})^{98}$Mo & 45$-$47 & 5, 12, 13, 35 \\
$^{96}$Zr$(\alpha,\mathrm{n})^{99}$Mo & 44$-$47 & 4, 5, 6, 7, 8, 12, 13, 16, 20, 21, 22, 23, 24, 25, 29, 30, 35 \\
$^{97}$Zr$(\alpha,\mathrm{n})^{100}$Mo & 44, 46, 47 & 4, 5, 6, 7, 8, 21, 22, 23, 24, 25, 29, 30 \\
$^{98}$Zr$(\alpha,\mathrm{n})^{101}$Mo & 44, 46, 47 & 6, 7, 8, 22, 23, 24, 25, 29, 30 \\
$^{97}$Nb$(\alpha,\mathrm{n})^{100}$Tc & 45, 46, 47 & 12, 13, 14, 15, 35 \\
  \hline
  \end{tabular}
 \end{center}
\end{table*}

\section{Conclusions}
\label{sec:conclusions}

Observations of the oldest stars in our galaxy and in dwarf galaxies point to an extra process contributing to the abundances of elements between Sr and Ag in addition to the s-process and r-process. It is possible that this extra process is a weak r-process that takes place in matter ejected by neutrinos in core-collapse supernovae  with only slightly neutron-rich conditions ($0.4<Y_e<0.5$). 
It is known that \alphan reaction are critical to move matter towards heavy elements in core-collapse supernovae \citep{Hoffman.etal:1996,Bliss.etal:2017}. Even though our understanding of the weak r-process has increased in the last years, the calculated abundances are still uncertain due to the lack of experimental information for \alphan reactions. 


In this paper, we have identified the most important reactions that need to be measured to reduce the nuclear physics uncertainty to be able to use observations to constrain astrophysical  wind  conditions.  We selected 36 tracers from \cite{Bliss.etal:2018} representing the evolution of ejected matter under a broad range of astrophysical  conditions. For each tracer, we have performed a Monte Carlo study varying over 900 \alphan reaction rates. In order to decide which reactions are most important, we used two criteria: an Spearman's correlation coefficient $|r_{\mathrm{corr}}|$ above $0.20$  and a significant impact on the abundance variation due to the reaction. Among the relevant reactions one can distinguish three groups depending on the nuclei involved. Reactions of nuclei at the shell closure $N=50$ have a clear impact on the final abundances. Since there is a $(n,\gamma)-(\gamma,n)$ equilibrium matter accumulates at $N=50$ leading to an enhanced importance of those nuclei. Reaction of nuclei with $N>50$ affect the abundances of heavier nuclei because those nuclei are reached  once the matter overcomes the shell closure. Reactions of nuclei with $N<50$ are less relevant. We provide a set of 45 \alphan reactions (Table~\ref{tab:complete}) that are relevant for the weak r-process in core-collapse supernovae. In addition to examining the correlation coefficient and the impact on the final abundances, we have checked the number of final elemental abundances that are affected by one reaction and the number of tracers in which a reaction is important. 


Future experiments will reduce the uncertainties of these reactions and will provide improved constraints to theoretical reaction models. This is critical to be able to combine nucleosynthesis calculations and observations  to understand the origin of lighter heavy elements and the astrophysical conditions where they are synthesised.  
\begin{acknowledgments}
This work was funded by  Deutsche Forschungsgemeinschaft through SFB 1245, ERC 677912 EUROPIUM, BMBF under grant No. 05P15RDFN1, and by the National Science Foundation under Grant No. PHY-1430152 (JINA Center for the Evolution of the Elements).  J.B. acknowledge the MGK of the SFB 1245 and the JINA Center for the Evolution of the Elements for support during a research stay at Michigan State University. 
\end{acknowledgments}

\appendix
\section{Tables}

\Cref{tab:Z36,tab:Z37,tab:Z38,tab:Z39,tab:Z40,tab:Z41,tab:Z42,tab:Z44,tab:Z45,tab:Z46,tab:Z47} show for the given element and for the given wind trajectories, the range of Spearman's correlation coefficients ($|r_{\mathrm{corr}}|$) and the abundance variation range within $2-\sigma$ of the average abundance in the Monte Carlo study (if multiple trajectories).

\begin{table*}[!htb]
 \begin{center}
  \caption{Z=36}
  \label{tab:Z36}
  \vspace*{3mm}
  \begin{tabular}{cccc}
  \hline
  \hline
Reaction  & Abundance variation & Correlation coefficient & MC tracers  \\
  \hline
$^{68}$Fe$(\alpha,\mathrm{n})$ & 8.38 & 0.35 & 3 \\
$^{71}$Co$(\alpha,\mathrm{n})$ & 8.38 & 0.26 & 3 \\
$^{74}$Ni$(\alpha,\mathrm{n})$ & 5.37-15.66 & 0.2-0.29 & 2, 3, 17, 18, 32 \\
$^{76}$Ni$(\alpha,\mathrm{n})$ & 7.77-15.66 & 0.22-0.3 & 2, 3, 18, 32 \\
$^{76}$Zn$(\alpha,\mathrm{n})$ & 5.37-15.66 & 0.2-0.32 & 2, 17, 18, 32 \\
$^{78}$Zn$(\alpha,\mathrm{n})$ & 5.37-15.66 & 0.25-0.3 & 2, 17, 18, 32 \\
$^{79}$Zn$(\alpha,\mathrm{n})$ & 7.77-15.66 & 0.3-0.32 & 2, 18, 32 \\
$^{80}$Zn$(\alpha,\mathrm{n})$ & 7.77-15.66 & 0.21-0.25 & 2, 18, 32 \\
$^{81}$Ga$(\alpha,\mathrm{n})$ & 5.37-7.77 & 0.23 & 17, 32 \\
$^{80}$Ge$(\alpha,\mathrm{n})$ & 12.33 & 0.22 & 28 \\
$^{82}$Ge$(\alpha,\mathrm{n})$ & 5.37-36.31 & 0.24-0.87 & 11, 17, 19, 27, 28 \\
$^{83}$As$(\alpha,\mathrm{n})$ & 12.33-27.43 & 0.21-0.39 & 27, 28 \\
$^{84}$Se$(\alpha,\mathrm{n})$ & 5.46-101.0 & 0.54-0.81 & 6, 7, 8, 9, 10, 22, 23, 24, 26, 29, 30, 31 \\
$^{85}$Se$(\alpha,\mathrm{n})$ & 5.46-101.0 & 0.32-0.53 & 6, 7, 8, 9, 10, 22, 23, 24, 26, 29, 30, 31 \\
  \hline
  \end{tabular}
 \end{center}
\end{table*}

\begin{table*}[!htb]
 \begin{center}
  \caption{Z=37}
  \label{tab:Z37}
  \vspace*{3mm}
  \begin{tabular}{cccc}
  \hline
  \hline
Reaction  & Abundance variation & Correlation coefficient & MC tracers  \\
  \hline
$^{68}$Fe$(\alpha,\mathrm{n})$ & 5.54 & 0.36 & 3 \\
$^{71}$Co$(\alpha,\mathrm{n})$ & 5.54 & 0.27 & 3 \\
$^{74}$Ni$(\alpha,\mathrm{n})$ & 5.54-16.18 & 0.26-0.28 & 2, 3, 18 \\
$^{76}$Ni$(\alpha,\mathrm{n})$ & 5.54-16.18 & 0.24-0.28 & 2, 3, 18 \\
$^{77}$Cu$(\alpha,\mathrm{n})$ & 5.54 & 0.2 & 3 \\
$^{76}$Zn$(\alpha,\mathrm{n})$ & 15.84-16.18 & 0.2-0.23 & 2, 18 \\
$^{78}$Zn$(\alpha,\mathrm{n})$ & 15.84-16.18 & 0.31 & 2, 18 \\
$^{79}$Zn$(\alpha,\mathrm{n})$ & 15.84-16.18 & 0.31-0.33 & 2, 18 \\
$^{80}$Zn$(\alpha,\mathrm{n})$ & 15.84-16.18 & 0.23-0.26 & 2, 18 \\
$^{80}$Ge$(\alpha,\mathrm{n})$ & 6.52 & 0.29 & 33 \\
$^{82}$Ge$(\alpha,\mathrm{n})$ & 6.52-23.15 & 0.55-0.69 & 17, 19, 33 \\
$^{83}$As$(\alpha,\mathrm{n})$ & 12.13-97.21 & 0.24-0.42 & 11, 26, 27, 28 \\
$^{84}$Se$(\alpha,\mathrm{n})$ & 5.07-97.21 & 0.21-0.6 & 9, 10, 11, 19, 22, 26, 27, 28, 31 \\
$^{85}$Se$(\alpha,\mathrm{n})$ & 8.49-97.21 & 0.2-0.49 & 6, 9, 10, 11, 26, 27, 28, 31 \\
$^{85}$Br$(\alpha,\mathrm{n})$ & 5.07-14.13 & 0.69-0.9 & 6, 7, 8, 22, 23, 24, 29, 30 \\
$^{87}$Br$(\alpha,\mathrm{n})$ & 8.49-11.72 & 0.21-0.22 & 6, 29 \\
$^{88}$Kr$(\alpha,\mathrm{n})$ & 35.33 & 0.21 & 32 \\
$^{90}$Kr$(\alpha,\mathrm{n})$ & 35.33-97.21 & 0.22-0.37 & 26, 32 \\
$^{94}$Sr$(\alpha,\mathrm{n})$ & 35.33-52.43 & 0.25-0.29 & 10, 32 \\
  \hline
  \end{tabular}
 \end{center}
\end{table*}

\begin{table*}[!htb]
 \begin{center}
  \caption{Z=38}
  \label{tab:Z38}
  \vspace*{3mm}
  \begin{tabular}{cccc}
  \hline
  \hline
Reaction  & Abundance variation & Correlation coefficient & MC tracers  \\
  \hline
$^{74}$Ni$(\alpha,\mathrm{n})$ & 11.87 & 0.28 & 3 \\
$^{76}$Ni$(\alpha,\mathrm{n})$ & 11.87 & 0.28 & 3 \\
$^{78}$Zn$(\alpha,\mathrm{n})$ & 10.16-84.76 & 0.2-0.26 & 2, 3, 18, 32 \\
$^{79}$Zn$(\alpha,\mathrm{n})$ & 10.16-84.76 & 0.21-0.27 & 2, 3, 18, 32 \\
$^{81}$Ga$(\alpha,\mathrm{n})$ & 10.16 & 0.2 & 32 \\
$^{80}$Ge$(\alpha,\mathrm{n})$ & 8.76 & 0.3 & 33 \\
$^{82}$Ge$(\alpha,\mathrm{n})$ & 8.76-32.1 & 0.53-0.64 & 17, 33 \\
$^{84}$Se$(\alpha,\mathrm{n})$ & 5.81-297.08 & 0.3-0.7 & 6, 7, 9, 10, 11, 19, 26, 27, 28, 29, 30, 31 \\
$^{85}$Se$(\alpha,\mathrm{n})$ & 7.57-297.08 & 0.21-0.53 & 6, 9, 10, 11, 19, 26, 27, 28, 29, 30, 31 \\
$^{85}$Br$(\alpha,\mathrm{n})$ & 10.85-21.93 & 0.43-0.46 & 9, 10, 31 \\
$^{86}$Kr$(\alpha,\mathrm{n})$ & 6.81-10.53 & 0.3-0.58 & 4, 5, 8, 16, 24, 25 \\
$^{87}$Kr$(\alpha,\mathrm{n})$ & 6.81-10.53 & 0.34-0.39 & 4, 5, 8, 16, 24, 25 \\
$^{88}$Kr$(\alpha,\mathrm{n})$ & 5.81-32.1 & 0.26-0.54 & 4, 5, 7, 8, 16, 17, 24, 25, 30 \\
$^{90}$Kr$(\alpha,\mathrm{n})$ & 11.87-297.08 & 0.28-0.34 & 3, 17, 19 \\
$^{92}$Sr$(\alpha,\mathrm{n})$ & 5.81 & 0.23 & 7 \\
$^{94}$Sr$(\alpha,\mathrm{n})$ & 5.81-84.76 & 0.26-0.45 & 2, 6, 7, 18, 29, 30 \\
  \hline
  \end{tabular}
 \end{center}
\end{table*}

\begin{table*}[!htb]
 \begin{center}
  \caption{Z=39}
  \label{tab:Z39}
  \vspace*{3mm}
  \begin{tabular}{cccc}
  \hline
  \hline
Reaction  & Abundance variation & Correlation coefficient & MC tracers  \\
  \hline
$^{59}$Fe$(\alpha,\mathrm{n})$ & 16.14 & 0.22 & 36 \\
$^{63}$Co$(\alpha,\mathrm{n})$ & 16.14 & 0.25 & 36 \\
$^{74}$Ni$(\alpha,\mathrm{n})$ & 13.44-14.08 & 0.2-0.3 & 3, 32 \\
$^{76}$Ni$(\alpha,\mathrm{n})$ & 13.44-14.08 & 0.2-0.31 & 3, 32 \\
$^{72}$Zn$(\alpha,\mathrm{n})$ & 16.14 & 0.29 & 36 \\
$^{76}$Zn$(\alpha,\mathrm{n})$ & 13.44 & 0.21 & 32 \\
$^{78}$Zn$(\alpha,\mathrm{n})$ & 13.44-107.03 & 0.23-0.28 & 2, 3, 18, 32 \\
$^{79}$Zn$(\alpha,\mathrm{n})$ & 13.44-107.03 & 0.24-0.29 & 2, 3, 18, 32 \\
$^{80}$Zn$(\alpha,\mathrm{n})$ & 37.57 & 0.22 & 18 \\
$^{81}$Ga$(\alpha,\mathrm{n})$ & 13.44 & 0.21 & 32 \\
$^{78}$Ge$(\alpha,\mathrm{n})$ & 16.14 & 0.24 & 36 \\
$^{80}$Ge$(\alpha,\mathrm{n})$ & 12.42 & 0.23 & 33 \\
$^{82}$Ge$(\alpha,\mathrm{n})$ & 12.42-248.76 & 0.35-0.47 & 17, 33 \\
$^{84}$Se$(\alpha,\mathrm{n})$ & 5.71-141.15 & 0.22-0.6 & 6, 9, 10, 11, 19, 26, 27, 28, 31, 36 \\
$^{85}$Se$(\alpha,\mathrm{n})$ & 8.36-141.15 & 0.25-0.51 & 9, 10, 11, 19, 26, 27, 28, 31 \\
$^{85}$Br$(\alpha,\mathrm{n})$ & 6.42-24.93 & 0.21-0.64 & 9, 10, 26, 28, 29, 31 \\
$^{87}$Br$(\alpha,\mathrm{n})$ & 8.36-12.0 & 0.22 & 9, 10, 31 \\
$^{88}$Br$(\alpha,\mathrm{n})$ & 10.57 & 0.21 & 26 \\
$^{86}$Kr$(\alpha,\mathrm{n})$ & 6.62 & 0.3 & 7 \\
$^{87}$Kr$(\alpha,\mathrm{n})$ & 6.62-7.51 & 0.24-0.33 & 7, 30 \\
$^{88}$Kr$(\alpha,\mathrm{n})$ & 5.71-248.76 & 0.26-0.51 & 6, 7, 17, 29, 30, 33 \\
$^{89}$Kr$(\alpha,\mathrm{n})$ & 5.71-248.76 & 0.21-0.27 & 6, 17, 29, 30 \\
$^{90}$Kr$(\alpha,\mathrm{n})$ & 5.71-248.76 & 0.2-0.39 & 6, 17, 19, 26, 28, 29, 30 \\
$^{94}$Sr$(\alpha,\mathrm{n})$ & 37.57-107.03 & 0.32-0.4 & 2, 18 \\
  \hline
  \end{tabular}
 \end{center}
\end{table*}

\begin{table*}[!htb]
 \begin{center}
  \caption{Z=40}
  \label{tab:Z40}
  \vspace*{3mm}
  \begin{tabular}{cccc}
  \hline
  \hline
Reaction  & Abundance variation & Correlation coefficient & MC tracers  \\
  \hline
$^{59}$Fe$(\alpha,\mathrm{n})$ & 11.26 & 0.22 & 36 \\
$^{63}$Co$(\alpha,\mathrm{n})$ & 6.0-11.26 & 0.21-0.23 & 34, 36 \\
$^{74}$Ni$(\alpha,\mathrm{n})$ & 12.76-18.57 & 0.25-0.29 & 3, 18 \\
$^{76}$Ni$(\alpha,\mathrm{n})$ & 12.76-18.57 & 0.26-0.3 & 3, 18 \\
$^{72}$Zn$(\alpha,\mathrm{n})$ & 11.26 & 0.27 & 36 \\
$^{76}$Zn$(\alpha,\mathrm{n})$ & 12.76 & 0.22 & 3 \\
$^{78}$Zn$(\alpha,\mathrm{n})$ & 12.76-50.35 & 0.24-0.3 & 2, 3, 18 \\
$^{79}$Zn$(\alpha,\mathrm{n})$ & 12.76-50.35 & 0.26-0.32 & 2, 3, 18 \\
$^{80}$Zn$(\alpha,\mathrm{n})$ & 12.76-18.57 & 0.22-0.25 & 3, 18 \\
$^{78}$Ge$(\alpha,\mathrm{n})$ & 11.26 & 0.27 & 36 \\
$^{84}$Se$(\alpha,\mathrm{n})$ & 11.26 & 0.2 & 36 \\
$^{86}$Kr$(\alpha,\mathrm{n})$ & 5.59-6.0 & 0.35-0.7 & 20, 34 \\
$^{87}$Kr$(\alpha,\mathrm{n})$ & 5.35-6.0 & 0.22-0.36 & 20, 28, 34 \\
$^{88}$Kr$(\alpha,\mathrm{n})$ & 5.04-6.0 & 0.24-0.44 & 11, 20, 27, 28, 34 \\
$^{89}$Kr$(\alpha,\mathrm{n})$ & 5.04-5.87 & 0.26-0.3 & 11, 27, 28 \\
$^{90}$Kr$(\alpha,\mathrm{n})$ & 5.04-5.87 & 0.4-0.55 & 11, 27, 28 \\
$^{94}$Sr$(\alpha,\mathrm{n})$ & 50.35 & 0.37 & 2 \\
  \hline
  \end{tabular}
 \end{center}
\end{table*}

\begin{table*}[!htb]
 \begin{center}
  \caption{Z=41}
  \label{tab:Z41}
  \vspace*{3mm}
  \begin{tabular}{cccc}
  \hline
  \hline
Reaction  & Abundance variation & Correlation coefficient & MC tracers  \\
  \hline
$^{59}$Fe$(\alpha,\mathrm{n})$ & 17.74 & 0.23 & 36 \\
$^{63}$Co$(\alpha,\mathrm{n})$ & 9.94-17.74 & 0.21-0.25 & 34, 36 \\
$^{74}$Ni$(\alpha,\mathrm{n})$ & 12.45-18.5 & 0.25-0.29 & 3, 18 \\
$^{76}$Ni$(\alpha,\mathrm{n})$ & 12.45-18.5 & 0.27-0.3 & 3, 18 \\
$^{72}$Zn$(\alpha,\mathrm{n})$ & 17.74 & 0.29 & 36 \\
$^{76}$Zn$(\alpha,\mathrm{n})$ & 11.25-12.45 & 0.21-0.22 & 3, 32 \\
$^{78}$Zn$(\alpha,\mathrm{n})$ & 11.25-33.1 & 0.27-0.31 & 2, 3, 18, 32 \\
$^{79}$Zn$(\alpha,\mathrm{n})$ & 11.25-33.1 & 0.27-0.32 & 2, 3, 18, 32 \\
$^{80}$Zn$(\alpha,\mathrm{n})$ & 12.45-33.1 & 0.2-0.25 & 2, 3, 18 \\
$^{81}$Ga$(\alpha,\mathrm{n})$ & 11.25 & 0.21 & 32 \\
$^{78}$Ge$(\alpha,\mathrm{n})$ & 17.74 & 0.28 & 36 \\
$^{82}$Ge$(\alpha,\mathrm{n})$ & 79.77 & 0.44 & 17 \\
$^{83}$As$(\alpha,\mathrm{n})$ & 5.99 & 0.27 & 33 \\
$^{84}$Se$(\alpha,\mathrm{n})$ & 5.22-21.78 & 0.21-0.51 & 11, 19, 27, 28, 29, 33 \\
$^{85}$Se$(\alpha,\mathrm{n})$ & 5.22-21.78 & 0.27-0.54 & 11, 19, 27, 28 \\
$^{86}$Kr$(\alpha,\mathrm{n})$ & 5.11-14.26 & 0.2-0.8 & 5, 13, 14, 15, 20, 33, 34 \\
$^{87}$Kr$(\alpha,\mathrm{n})$ & 5.11-9.94 & 0.22-0.35 & 5, 13, 20, 33, 34 \\
$^{88}$Kr$(\alpha,\mathrm{n})$ & 5.11-79.77 & 0.21-0.34 & 5, 11, 13, 17, 20, 27, 28, 29, 33, 34 \\
$^{90}$Kr$(\alpha,\mathrm{n})$ & 5.22-79.77 & 0.23-0.31 & 11, 17, 27, 28, 29 \\
$^{87}$Rb$(\alpha,\mathrm{n})$ & 5.48-14.26 & 0.23-0.64 & 14, 15 \\
$^{89}$Rb$(\alpha,\mathrm{n})$ & 5.11-5.98 & 0.22-0.29 & 5, 13, 20 \\
$^{94}$Sr$(\alpha,\mathrm{n})$ & 6.78-33.1 & 0.25-0.44 & 2, 11, 27, 29 \\
  \hline
  \end{tabular}
 \end{center}
\end{table*}

\begin{table*}[!htb]
 \begin{center}
  \caption{Z=42}
  \label{tab:Z42}
  \vspace*{3mm}
  \begin{tabular}{cccc}
  \hline
  \hline
Reaction  & Abundance variation & Correlation coefficient & MC tracers  \\
  \hline
$^{59}$Fe$(\alpha,\mathrm{n})$ & 8.55-156.19 & 0.2-0.23 & 34, 36 \\
$^{63}$Co$(\alpha,\mathrm{n})$ & 8.55-156.19 & 0.22-0.26 & 34, 36 \\
$^{74}$Ni$(\alpha,\mathrm{n})$ & 13.63 & 0.25 & 3 \\
$^{76}$Ni$(\alpha,\mathrm{n})$ & 13.63 & 0.27 & 3 \\
$^{72}$Zn$(\alpha,\mathrm{n})$ & 156.19 & 0.31 & 36 \\
$^{76}$Zn$(\alpha,\mathrm{n})$ & 13.63 & 0.23 & 3 \\
$^{78}$Zn$(\alpha,\mathrm{n})$ & 13.63 & 0.31 & 3 \\
$^{79}$Zn$(\alpha,\mathrm{n})$ & 13.63 & 0.32 & 3 \\
$^{80}$Zn$(\alpha,\mathrm{n})$ & 13.63 & 0.24 & 3 \\
$^{78}$Ge$(\alpha,\mathrm{n})$ & 156.19 & 0.29 & 36 \\
$^{80}$Ge$(\alpha,\mathrm{n})$ & 156.19 & 0.2 & 36 \\
$^{84}$Se$(\alpha,\mathrm{n})$ & 5.05-104.93 & 0.31-0.32 & 18, 20 \\
$^{85}$Se$(\alpha,\mathrm{n})$ & 104.93 & 0.3 & 18 \\
$^{86}$Kr$(\alpha,\mathrm{n})$ & 5.05-8.55 & 0.22-0.54 & 7, 14, 15, 20, 34 \\
$^{87}$Kr$(\alpha,\mathrm{n})$ & 5.6-8.55 & 0.25-0.26 & 7, 34 \\
$^{88}$Kr$(\alpha,\mathrm{n})$ & 5.6-8.55 & 0.29-0.37 & 6, 7, 19, 22, 29, 30, 34 \\
$^{89}$Kr$(\alpha,\mathrm{n})$ & 7.02-7.42 & 0.2-0.3 & 6, 19 \\
$^{90}$Kr$(\alpha,\mathrm{n})$ & 5.94-8.08 & 0.23-0.57 & 6, 19, 22, 26, 29 \\
$^{89}$Rb$(\alpha,\mathrm{n})$ & 5.05-8.55 & 0.21-0.27 & 7, 20, 34 \\
$^{88}$Sr$(\alpha,\mathrm{n})$ & 5.35-8.13 & 0.4-0.43 & 14, 15 \\
$^{89}$Sr$(\alpha,\mathrm{n})$ & 5.35-8.13 & 0.25-0.3 & 14, 15 \\
$^{90}$Sr$(\alpha,\mathrm{n})$ & 5.35-8.13 & 0.29-0.43 & 14, 15 \\
$^{92}$Sr$(\alpha,\mathrm{n})$ & 11.02 & 0.2 & 31 \\
$^{93}$Sr$(\alpha,\mathrm{n})$ & 8.67-11.02 & 0.24-0.25 & 9, 10, 31 \\
$^{94}$Sr$(\alpha,\mathrm{n})$ & 5.6-104.93 & 0.24-0.6 & 2, 6, 7, 9, 10, 18, 22, 26, 29, 30, 31 \\
  \hline
  \end{tabular}
 \end{center}
\end{table*}

\begin{table*}[!htb]
 \begin{center}
  \caption{Z=44}
  \label{tab:Z44}
  \vspace*{3mm}
  \begin{tabular}{cccc}
  \hline
  \hline
Reaction  & Abundance variation & Correlation coefficient & MC tracers  \\
  \hline
$^{84}$Se$(\alpha,\mathrm{n})$ & 5.61-24.77 & 0.22-0.31 & 2, 6, 18, 20, 22, 34 \\
$^{85}$Se$(\alpha,\mathrm{n})$ & 5.61-5.81 & 0.28 & 2, 18 \\
$^{86}$Kr$(\alpha,\mathrm{n})$ & 7.71-31.16 & 0.5-0.54 & 14, 15 \\
$^{88}$Kr$(\alpha,\mathrm{n})$ & 5.61-23.87 & 0.2-0.27 & 2, 8, 9, 18, 26, 31 \\
$^{89}$Kr$(\alpha,\mathrm{n})$ & 5.61-6.67 & 0.2-0.24 & 2, 18, 26 \\
$^{90}$Kr$(\alpha,\mathrm{n})$ & 5.61-10.05 & 0.22-0.47 & 2, 9, 10, 18, 26, 31 \\
$^{88}$Sr$(\alpha,\mathrm{n})$ & 7.71 & 0.22 & 15 \\
$^{90}$Sr$(\alpha,\mathrm{n})$ & 8.32-16.83 & 0.21-0.43 & 4, 5, 12, 13, 16, 20, 35 \\
$^{91}$Sr$(\alpha,\mathrm{n})$ & 8.32-9.02 & 0.2-0.21 & 5, 12, 13 \\
$^{92}$Sr$(\alpha,\mathrm{n})$ & 5.84-25.17 & 0.2-0.42 & 4, 5, 6, 7, 8, 12, 13, 16, 20, 21, 22, 23, 24, 25, 29, 30, 31, 34 \\
$^{93}$Sr$(\alpha,\mathrm{n})$ & 5.84-22.0 & 0.21-0.29 & 6, 7, 9, 10, 22, 29, 30, 31 \\
$^{94}$Sr$(\alpha,\mathrm{n})$ & 5.81-22.0 & 0.23-0.62 & 2, 6, 7, 9, 10, 22, 23, 26, 29, 30, 31 \\
$^{94}$Zr$(\alpha,\mathrm{n})$ & 7.71-31.16 & 0.32-0.4 & 14, 15 \\
$^{96}$Zr$(\alpha,\mathrm{n})$ & 8.32-25.17 & 0.22-0.4 & 4, 5, 8, 12, 13, 16, 21, 23, 24, 25, 35 \\
$^{97}$Zr$(\alpha,\mathrm{n})$ & 20.09-25.17 & 0.23-0.25 & 8, 23, 24 \\
$^{98}$Zr$(\alpha,\mathrm{n})$ & 20.09-25.17 & 0.2-0.22 & 8, 23, 24 \\
  \hline
  \end{tabular}
 \end{center}
\end{table*}

\begin{table*}[!htb]
 \begin{center}
  \caption{Z=45}
  \label{tab:Z45}
  \vspace*{3mm}
  \begin{tabular}{cccc}
  \hline
  \hline
Reaction  & Abundance variation & Correlation coefficient & MC tracers  \\
  \hline
$^{59}$Fe$(\alpha,\mathrm{n})$ & 61.36 & 0.25 & 34 \\
$^{63}$Co$(\alpha,\mathrm{n})$ & 50.02-61.36 & 0.21-0.27 & 20, 34 \\
$^{84}$Se$(\alpha,\mathrm{n})$ & 5.98-50.02 & 0.29-0.36 & 2, 20 \\
$^{85}$Se$(\alpha,\mathrm{n})$ & 5.44-5.98 & 0.22-0.3 & 2, 18 \\
$^{86}$Kr$(\alpha,\mathrm{n})$ & 10.69-31.69 & 0.22-0.56 & 4, 5, 13, 14, 15, 35 \\
$^{87}$Kr$(\alpha,\mathrm{n})$ & 21.89 & 0.25 & 5 \\
$^{88}$Kr$(\alpha,\mathrm{n})$ & 5.23-29.37 & 0.21-0.33 & 5, 6, 7, 11, 18, 19, 22, 27, 30, 32 \\
$^{89}$Kr$(\alpha,\mathrm{n})$ & 5.23-8.43 & 0.23-0.27 & 11, 18, 19, 27, 32 \\
$^{90}$Kr$(\alpha,\mathrm{n})$ & 5.23-8.43 & 0.4-0.53 & 2, 11, 18, 19, 27, 32 \\
$^{87}$Rb$(\alpha,\mathrm{n})$ & 31.69 & 0.22 & 14 \\
$^{90}$Sr$(\alpha,\mathrm{n})$ & 8.51-50.02 & 0.2-0.35 & 4, 5, 12, 13, 16, 20, 35 \\
$^{91}$Sr$(\alpha,\mathrm{n})$ & 8.51 & 0.21 & 16 \\
$^{92}$Sr$(\alpha,\mathrm{n})$ & 8.39-61.36 & 0.2-0.39 & 4, 5, 6, 7, 8, 13, 16, 21, 22, 23, 24, 25, 29, 30, 34 \\
$^{93}$Sr$(\alpha,\mathrm{n})$ & 21.65-29.37 & 0.22-0.27 & 6, 7, 29, 30 \\
$^{94}$Sr$(\alpha,\mathrm{n})$ & 5.23-29.37 & 0.21-0.62 & 2, 6, 7, 8, 11, 19, 22, 23, 24, 25, 27, 29, 30, 32 \\
$^{94}$Y$(\alpha,\mathrm{n})$ & 8.39-12.04 & 0.21-0.31 & 4, 8, 16, 21, 23, 24, 25 \\
$^{95}$Y$(\alpha,\mathrm{n})$ & 8.39-12.04 & 0.22-0.3 & 8, 23, 24, 25 \\
$^{94}$Zr$(\alpha,\mathrm{n})$ & 10.69-14.08 & 0.2-0.34 & 15, 35 \\
$^{95}$Zr$(\alpha,\mathrm{n})$ & 10.89-14.08 & 0.21-0.23 & 12, 35 \\
$^{96}$Zr$(\alpha,\mathrm{n})$ & 10.89-14.91 & 0.32-0.46 & 12, 13, 35 \\
$^{97}$Nb$(\alpha,\mathrm{n})$ & 10.69-31.69 & 0.23-0.53 & 14, 15, 35 \\
  \hline
  \end{tabular}
 \end{center}
\end{table*}

\begin{table*}[!htb]
 \begin{center}
  \caption{Z=46}
  \label{tab:Z46}
  \vspace*{3mm}
  \begin{tabular}{cccc}
  \hline
  \hline
Reaction  & Abundance variation & Correlation coefficient & MC tracers  \\
  \hline
$^{90}$Kr$(\alpha,\mathrm{n})$ & 6.53 & 0.2 & 27 \\
$^{90}$Sr$(\alpha,\mathrm{n})$ & 15.31-22.98 & 0.23-0.27 & 12, 13, 16, 20, 35 \\
$^{92}$Sr$(\alpha,\mathrm{n})$ & 5.9-22.15 & 0.2-0.27 & 7, 16, 28, 29, 30, 31 \\
$^{93}$Sr$(\alpha,\mathrm{n})$ & 5.9-16.65 & 0.21-0.26 & 6, 9, 10, 26, 27, 28, 29, 31 \\
$^{94}$Sr$(\alpha,\mathrm{n})$ & 5.9-16.65 & 0.29-0.66 & 6, 7, 9, 10, 22, 26, 27, 28, 29, 30, 31 \\
$^{95}$Y$(\alpha,\mathrm{n})$ & 9.36 & 0.21 & 30 \\
$^{95}$Zr$(\alpha,\mathrm{n})$ & 15.31-20.4 & 0.2-0.23 & 5, 12, 13 \\
$^{96}$Zr$(\alpha,\mathrm{n})$ & 8.7-42.62 & 0.22-0.62 & 4, 5, 7, 8, 12, 13, 16, 20, 21, 23, 24, 25, 30, 35 \\
$^{97}$Zr$(\alpha,\mathrm{n})$ & 20.4-42.62 & 0.2-0.3 & 4, 5, 8, 21, 23, 24, 25 \\
$^{98}$Zr$(\alpha,\mathrm{n})$ & 23.65-28.26 & 0.2-0.3 & 8, 23, 24, 25 \\
$^{97}$Nb$(\alpha,\mathrm{n})$ & 15.31-17.58 & 0.21-0.37 & 12, 13, 35 \\
  \hline
  \end{tabular}
 \end{center}
\end{table*}

\begin{table*}[!htb]
 \begin{center}
  \caption{Z=47}
  \label{tab:Z47}
  \vspace*{3mm}
  \begin{tabular}{cccc}
  \hline
  \hline
Reaction  & Abundance variation & Correlation coefficient & MC tracers  \\
  \hline
$^{67}$Cu$(\alpha,\mathrm{n})$ & 24.85 & 0.23 & 35 \\
$^{86}$Kr$(\alpha,\mathrm{n})$ & 20.78 & 0.22 & 13 \\
$^{90}$Sr$(\alpha,\mathrm{n})$ & 17.92-27.91 & 0.22-0.23 & 12, 16 \\
$^{92}$Sr$(\alpha,\mathrm{n})$ & 8.2-9.1 & 0.21-0.26 & 11, 28, 31 \\
$^{93}$Sr$(\alpha,\mathrm{n})$ & 7.61-11.35 & 0.22-0.25 & 9, 11, 26, 27, 28, 31 \\
$^{94}$Sr$(\alpha,\mathrm{n})$ & 7.61-29.91 & 0.23-0.63 & 6, 9, 10, 11, 26, 27, 28, 29, 31 \\
$^{95}$Zr$(\alpha,\mathrm{n})$ & 17.92 & 0.24 & 12 \\
$^{96}$Zr$(\alpha,\mathrm{n})$ & 17.92-42.23 & 0.2-0.6 & 4, 5, 6, 7, 8, 12, 13, 16, 21, 22, 23, 24, 25, 29, 30, 35 \\
$^{97}$Zr$(\alpha,\mathrm{n})$ & 18.74-42.23 & 0.24-0.33 & 4, 6, 7, 8, 21, 22, 23, 24, 25, 29, 30 \\
$^{98}$Zr$(\alpha,\mathrm{n})$ & 18.74-42.23 & 0.2-0.35 & 6, 7, 8, 22, 23, 24, 25, 29, 30 \\
$^{97}$Nb$(\alpha,\mathrm{n})$ & 17.92-24.85 & 0.2-0.32 & 12, 13, 35 \\
  \hline
  \end{tabular}
 \end{center}
\end{table*}



\bibliography{paper}

\end{document}